\documentclass{article}
\usepackage{spconf,amsmath,graphicx,hyperref, float}

\usepackage[capitalize]{cleveref}
\usepackage{booktabs}

\title{Benchmarking Diarization Models}
\name{Luca A. Lanzendörfer \qquad Florian Grötschla \qquad Cesare Blaser \qquad Roger Wattenhofer}
\address{ETH Zurich}

\begin{document}
\maketitle
\begin{abstract}
Speaker diarization is the task of partitioning audio into segments according to speaker identity, answering the question of ``who spoke when'' in multi-speaker conversation recordings. While diarization is an essential task for many downstream applications, it remains an unsolved problem. Errors in diarization propagate to downstream systems and cause wide-ranging failures. To this end, we examine exact failure modes by evaluating five state-of-the-art diarization models, across four diarization datasets spanning multiple languages and acoustic conditions. The evaluation datasets consist of 196.6 hours of multilingual audio, including English, Mandarin, German, Japanese, and Spanish. Overall, we find that PyannoteAI achieves the best performance at 11.2\% DER, while DiariZen provides a competitive open-source alternative at 13.3\% DER. When analyzing failure cases, we find that the primary cause of diarization errors stem from missed speech segments followed by speaker confusion, especially in high-speaker count settings. 
\end{abstract}
\section{Introduction}
Speaker diarization, the task of finding time stamp and speaker ID information for spoken conversations, remains one of the most challenging problems in speech processing. The difficulty arises from overlapping speech segments, varying numbers of speakers, and heterogeneous recording environments \cite{park2022review}. Accurate speaker diarization has become increasingly critical for automated meeting transcriptions, voice chat analysis, broadcast content processing, and as preprocessing for automatic speech recognition \cite{park2022review}.

Contemporary diarization systems follow three primary paradigms. Modular approaches decompose the task into specialized subtasks by first detecting speech through voice activity detection (VAD), followed by segmentation, embedding extraction, and speaker clustering~\cite{park2022review,landini2020vbx, bredin2019pyannote}. Various approaches additionally introduce overlap-aware segmentation~\cite{plaquet2023pyannote, bredin2021EEsegm}. Although this modularity enables plug-and-play replacement, error propagation from previous modules (e.g., errors in VAD) are a significant drawback of these pipelines. End-to-end neural diarization (EEND) formulates the diarization problem as a multi-label sequence classification with joint optimization~\cite{fujita2019EEND,medennikov2020TSVAD,park2025sortformer}, allowing explicit overlap modeling but facing scalability challenges with high speaker counts and long recordings. Recent EEND developments include streaming capabilities through mechanisms such as the Arrival-Order Speaker Cache~\cite{medennikov2025streamingsortformerspeakercachebased}. Hybrid approaches combine these paradigms, applying local EEND processing on audio chunks with global clustering~\cite{kinoshita2021BobW-rev}, balancing overlap handling with scalability. 

Recent benchmarking studies~\cite{pacheco2025sdbench,aperdannier2024diareval} have provided comparative insights into diarization system performance. However, comprehensive evaluation across diverse languages and recording settings using readily accessible models remains limited.
To this end, we evaluate five diarization systems across four datasets spanning five languages and recording conditions (i.e., meeting recordings, phone calls, and in-the-wild audio). The tested diarization models consist of four open-source state-of-the-art models representing different paradigms, and one commercial API. 

This work presents a comparative evaluation of speaker diarization models. Our evaluation addresses two questions. Firstly, the practical question of ``given an audio recording in a specific language and acoustic condition, which available model achieves the best diarization performance?'' and secondly, with future improvements in mind: What are the exact causes of failure for current state-of-the-art diarization models? By evaluating the performance of these models through multilingual analysis, speaker scalability, and computational efficiency, we provide users with practical guidance across different deployment scenarios, and highlight specific areas of improvement for the next generation of diarization models.

\section{Methodology}
Our evaluation adopts an out-of-the-box methodology, testing each system as documented without parameter tuning or modifications, providing a baseline assessment of plug-and-play deployment performance. All audio was standardized to mono-channel 16 kHz sampling rate for consistent input across models. All experiments were conducted on our internal cluster using NVIDIA RTX A6000 GPUs except for pyannotAI, where we used the provided cloud-based API.

\subsection{Models}
\label{ssec:models}

\noindent \textbf{Pyannote.} A widely-used open-source modular diarization pipeline~\cite{bredin2023pyannote, plaquet2023pyannote}. The system performs diarization by chunking audio into overlapping segments, applying ResNet-based neural segmentation to detect speech regions, extracting speaker embeddings, and clustering them to assign speaker identities. We use the 3.1 version provided on Hugging Face.\footnote{https://huggingface.co/pyannote/speaker-diarization-3.1}
PyannoteAI\footnote{https://www.pyannote.ai} is a cloud-based commercial service based on the Pyannote ``precision-2'' model, although its exact details of implementation are not publicly documented. %

\smallskip
\noindent \textbf{Sortformer (SF).} An end-to-end neural diarization model~\cite{park2025sortformer} using a transformer architecture with arrival time sorting, and optimized for up to four speakers. The authors mention that the model achieves degraded performance for higher speaker counts and that the model suffers from computational limitations for long-form audio processing, due to the quadratic memory scaling of the self-attention mechanism. We use the Hugging Face version in our experiments.\footnote{https://huggingface.co/nvidia/diar\_sortformer\_4spk-v1}

\begin{table}[t!]
\centering
\begin{tabular}{lrrrr}
\toprule
\textbf{\#Speakers} & \textbf{CH} & \textbf{VC} & \textbf{AMI} & \textbf{ALI} \\
\midrule
1   & 0  & 3.2   & 0 & 0 \\         
2   & 74.2 & 7.9   & 0 & 5.9 \\         
3   & 16.2 & 4.4   & 0.8 & 2.7 \\         
4   & 6.0  & 4.7   & 17.9 & 6.4 \\         
5+  & 2.6  & 43.6 & 0 & 0 \\   
\bottomrule
\end{tabular}
\caption{Speaker count distribution (in hours) by dataset. CALLHOME (CH) contains mostly two-speaker phone conversations. VoxConverse (VC) has 43.6 hours of audio with more than 5 speakers, with some samples reaching 15+ active speakers, coming from TV program recordings and public press events. AMI and ALI contain mostly meeting recordings with 4 active speakers.}
\label{tab:spkcount}
\end{table}

\smallskip
\noindent \textbf{Sortformer v2 (SF v2).} A recent improvement over Sortformer, Sortformer v2~\cite{medennikov2025streamingsortformerspeakercachebased} consists of an end-to-end neural diarization model featuring Arrival-Order Speaker Cache (AOSC), designed for improved long-form audio processing capabilities. AOSC is a fixed-length buffer that stores per-speaker embeddings in arrival order while retaining the most recent and high-confidence frames, which helps resolve between-chunk speaker permutation automatically and enables robust low-latency streaming diarization.\footnote{https://huggingface.co/nvidia/diar\_streaming\_sortformer\_4spk-v2}

\smallskip
\noindent \textbf{DiariZen.} A hybrid diarization approach~\cite{han2024diarizen, han2025fine, han2025efficient}, combining end-to-end neural diarization processing with WavLM~\cite{chen2022wavlm} for speaker embeddings and clustering components from Pyannote.\footnote{https://huggingface.co/BUT-FIT/diarizen-wavlm-large-s80-md}

\subsection{Model Adaptations \& Evaluation Metric}

\textbf{Sortformer chunking.} Sortformer (SF) requires chunking for recordings exceeding 12 minutes (specific to the hardware we used and as outlined by the authors), due to memory scaling constraints in its attention-based architecture. We implemented audio chunking by splitting recordings into 12-minute chunks, with boundaries aligned with silence gaps predicted with SileroVAD~\cite{SileroVAD}, in order to preserve speech continuity and conversational dynamics. Only AMI and ALI datasets required chunking, which maintained stable speaker count and overlap characteristics with minimal deviation (0.5\%). 

\smallskip
\noindent \textbf{Sortformer v2 adaptations.} Sortformer v2 (SF v2) is evaluated on both the chunked audio that was processed for use by Sortformer, and the full original audio used by other models. We denote the model evaluated on full-length audio as Sortformer v2-streaming (SF v2-stream).

After processing, all model outputs were parsed into a standardized representation consisting of temporal segments with start times, end times, and speaker identifiers, allowing for consistent downstream evaluation.

\smallskip
\noindent \textbf{Diarization error rate (DER) as evaluation metric.}
We evaluate all models using DER, the standard metric for speaker diarization performance~\cite{pacheco2025sdbench}, quantifying the fraction of time attributed to incorrect labels. DER decomposes into three error components: missed speech (failure to detect actual speech regions), false alarm (detection of speech in non-speech regions), and speaker confusion (incorrect speaker assignment). We implement DER using pyannote.metrics\footnote{https://github.com/pyannote/pyannote-metrics} with a 0.25 second collar and \texttt{skip\_overlap=False} to evaluate performance on overlapping speech regions where diarization systems typically struggle most.

\begin{figure*}[t!]
    \centering
    \includegraphics[width=1\linewidth]{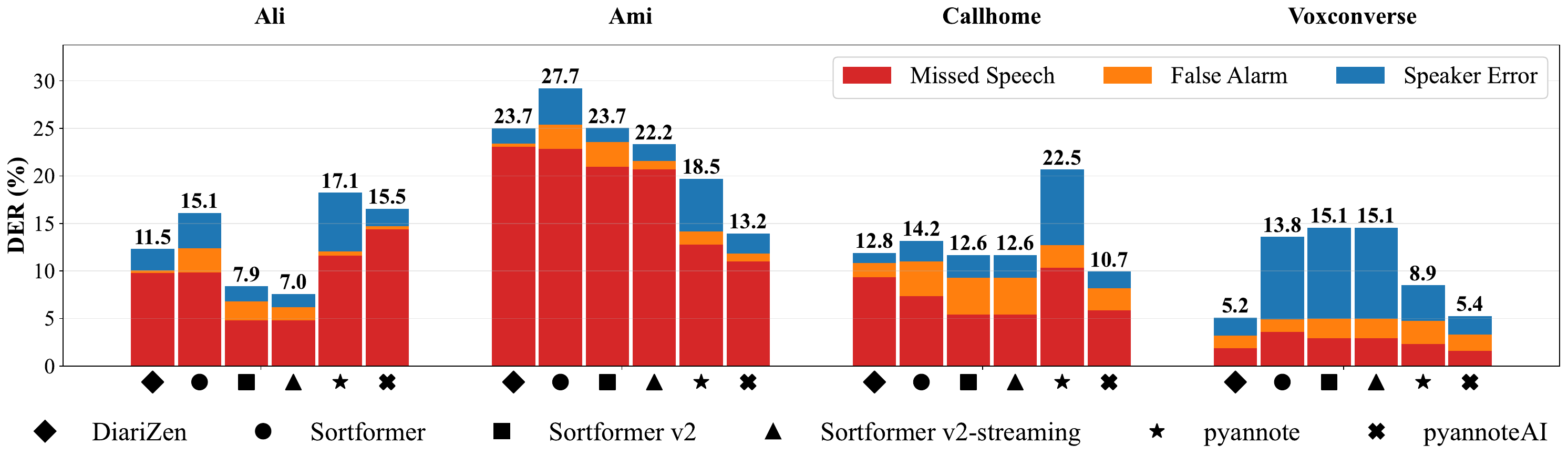}
    \caption{Diarization error rate across models showing missed speech (red), false alarm (orange), and speaker confusion (blue) components. We find that Sortformer v2-streaming, DiariZen, and the commercial PyannoteAI models perform the best overall. The numbers above the bar represent the total DER (\%, lower is better) for each model. We find that missed speech errors are the most common source of errors, with the exception of Sortformer, which has a higher speaker confusion error than other diarization models.}
    \label{fig:der_brakedown}
\end{figure*}

\subsection{Datasets}
\label{ssec:datasets}

\textbf{CALLHOME Corpus (CH).} A widely used multilingual telephone conversation dataset\footnote{https://huggingface.co/datasets/talkbank/callhome} spanning five languages: English (20.3h), Mandarin (20.3h), Japanese (18.7h), German (18.4h), and Spanish (21.3h). CH contains two speakers per session, and an average of 12.58\% overlapped speech (with 7.31 standard deviation). 

\smallskip
\noindent \textbf{VoxConverse v0.3 (VC).} An English language conversation dataset~\cite{chung2020VoxConverse} containing 63.8 hours of speech, where samples were extracted from YouTube videos containing between 2 and more than 20 speakers per recording. VoxConverse is widely used for high speaker count evaluation and in-the-wild acoustic conditions including background noise, varying microphone distances, and uncontrolled recording environments. VoxConverse contains an average 3.52\% overlap speech (with 5.71 standard deviation). We evaluated the models on both the ``dev'' and ``test'' split.

\smallskip
\noindent \textbf{AMI Meeting Corpus (AMI)}. A dataset~\cite{carletta2007AMI} containing English language scenario-based business meetings with exactly 4 participants per session (18.7 hours total). Used as a widely adopted benchmark and because professional meeting environments constitute a core application domain for speaker diarization. AMI contains on avarage 15.94\% of overlap speech (with 6.55 standard deviation). We evaluated the models on both ``dev'' and ``test'' split.

\smallskip
\noindent \textbf{AliMeeting (ALI).} Dataset~\cite{Yu2022M2MeT} contains 15 hours of Mandarin meeting scenarios with structure similar to AMI, featuring high speech overlap averaging 19\% (17.56 standard deviation), with some samples exceeding 50\%. We use ALI to evaluate Mandarin in addition to CALLHOME and challenging overlap conditions.
We evaluated the models on both ``eval'' and ``test'' split.

\begin{figure}[t!]
    \centering
    \includegraphics[width=1\linewidth]{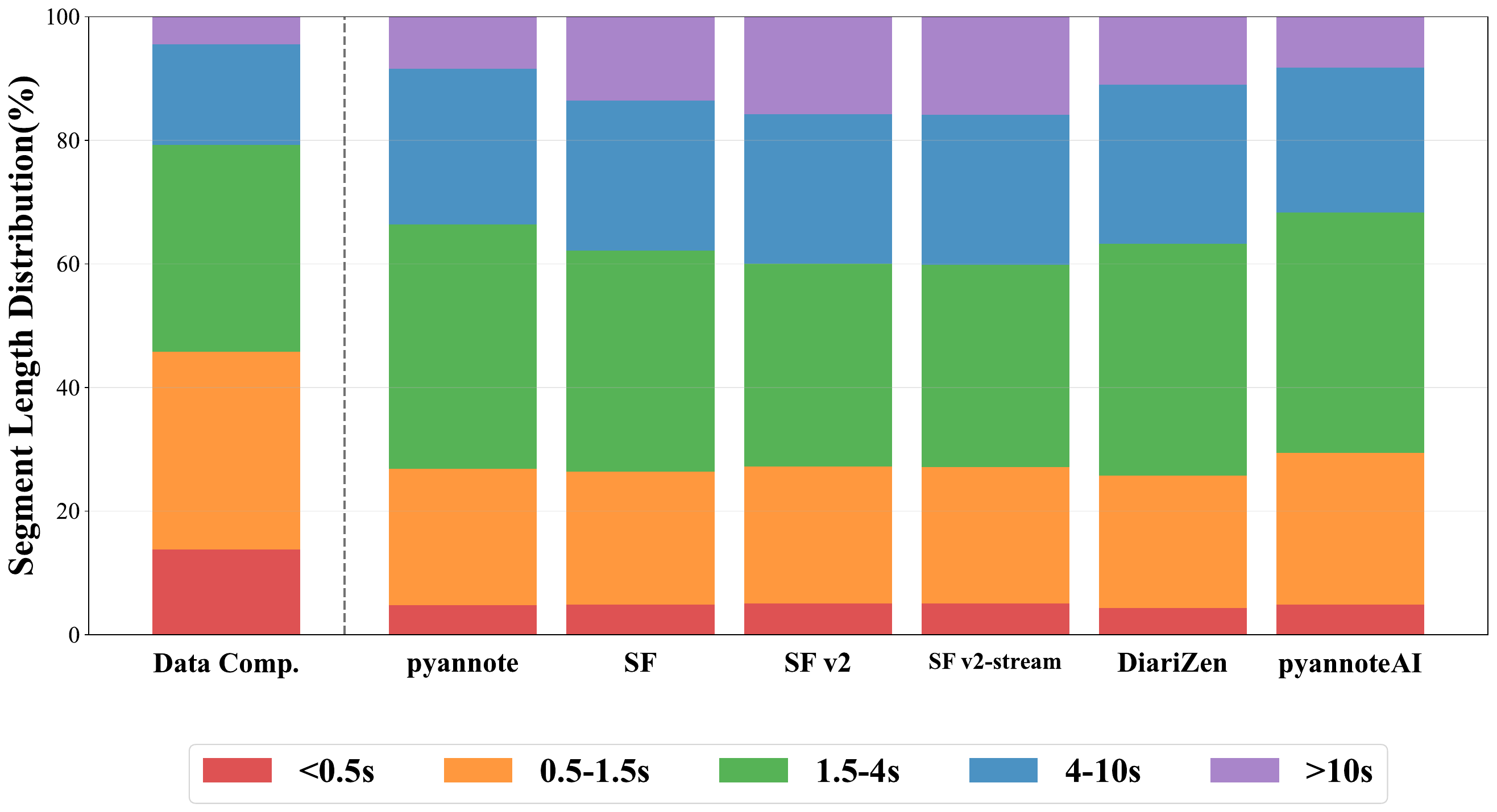}
    \caption{We visualize the distribution of speech segment lengths found over all datasets (left). For each diarization model we visualize in which segment length missed speech occurred (right). We find that models exhibit mostly unbiased error patterns across segment lengths, matching dataset composition, indicating that the failures are not from missing short segments but rather from inaccurate onset and end timestamp information across all segment length types.}
    \label{fig:error_related_gt_length}
\end{figure}

\section{Results}

\subsection{General Performance}
\cref{fig:der_brakedown} presents DER and its components across all model-dataset combinations, revealing that pyannoteAI achieves the best overall performance (DER = 11.2\%). DiariZen produces competitive performance (DER = 13.3\%) as well, with particularly strong results on VoxConverse (DER = 5.2\%), while Sortformer v2-streaming demonstrates exceptional performance on ALI (DER = 7.0\%), demonstrating that certain diarization models seem to be more suited for different conversational settings.

In terms of computational efficiency (cf.~\cref{tab:rtf}), we find that Sortformer's architecture delivers exceptional speed (regardless of the version), with SF v2 achieving the fastest average processing (RTF = 214.3x) across all datasets, also enabling real-time deployment through streaming~\cite{medennikov2025streamingsortformerspeakercachebased}. This represents a significant advantage for latency-critical applications, while still maintaining competitive diarization error rates.

Analyzing diarization failure modes (cf.~\cref{fig:der_brakedown}), we find that missed speech constitutes the dominant failure case across all diarization models, especially in meeting scenarios such as those covered in AMI and ALI. Additionally, we find that on VoxConverse all Sortformer versions exhibit increased speaker confusion errors. As VC contains on average higher average speaker counts (cf.~\cref{tab:spkcount}) this reflects the documented 4-speaker limit and highlights out-of-distribution generalization gaps.

Additionally, we analyze missed speech segments to understand error characteristics. All models show consistent average missed segment durations of roughly 350ms. Analysis of ground truth segments contributing to these errors (cf.~\cref{fig:error_related_gt_length}) reveals that models exhibit minimal bias toward specific segment lengths, with the exception for very short segments ($<0.5$s), which account for less than 5\% of errors. The brief duration of missed speech segments predominantly combined with the distribution highlighted in \cref{fig:error_related_gt_length} shows that the failure is in boundary precision of longer segments, rather than completely failing to detect shorter segments. This finding suggests that focusing on the precision of speech onset and end timestamps would significantly improve diarization performance.

\subsection{Per-Language Performance}
\cref{tab:der_by_lang} shows cross-lingual performance patterns across five languages. PyannoteAI achieves the best performance on English (DER = 6.6\%), German (DER = 8.3\%), and Spanish (DER = 14.3\%), while SF v2 and SF v2-stream both excel on Mandarin (DER = 9.2\%) and Japanese (DER = 12.7\%), suggesting different models have distinct cross-lingual strengths. English and Mandarin appear to be generally the languages where diarization performs the best, likely due to more training data availability, while Spanish presents the most challenging condition out of the evaluated languages.

\subsection{Per-Speaker Count Performance}
\cref{tab:der_by_spk} reveals speaker count scalability patterns. PyannoteAI demonstrates the best scores in terms of scalability, maintaining relatively stable performance from 2 speakers (DER = 9.9\%) to 5+ speakers (DER = 6.6\%). SF v2 and SF v2-stream show improved speaker handling compared to SF, with SF v2-stream exhibiting better performance on 4-speaker scenarios (DER = 13.2\% compared with SF's DER = 21.3\%), though all Sortformer-based models still experience degradation with increasing speaker counts due to their 4-speaker limit. DiariZen's hybrid approach demonstrates particular effectiveness in high-speaker scenarios (5+: DER = 7.1\%), with its strategy of processing local audio chunks with EEND from WavLM~\cite{chen2022wavlm}.

\begin{table}
\centering
\begin{tabular}{lccccc}
\toprule
Model & Zho & Eng & Deu & Jpn & Spa \\
\midrule
DiariZen        & 10.1 & 7.0 & 11.6 & 15.6 & 19.1 \\
SF               & 13.1 & 15.5 & 11.1 & 16.5 & 21.9 \\
SF v2           & \textbf{9.2} & 15.3 & 9.6 & \textbf{12.7} & 21.1 \\
SF v2-stream        & 9.4 & 14.1 & 9.6 & \textbf{12.7} & 21.1 \\
pyannote        & 19.8 & 11.5 & 19.0 & 28.8 & 27.3 \\
pyannoteAI      & 10.0 & \textbf{6.6} & \textbf{8.3} & 13.8 & \textbf{14.3} \\
\bottomrule
\end{tabular}
\caption{DER (\%, lower is better) aggregated by language. SF stands for Sortformer, SF v2 for Sortformer v2 with chunked audio, and SF v2-stream for Sortformer v2-streaming.}
\label{tab:der_by_lang}
\end{table}

\begin{table}
\centering
\begin{tabular}{lccccc}
\toprule
Model & 1 spk & 2 spk & 3 spk & 4 spk & 5+ spk \\
\midrule
DiariZen    & 2.3 & 11.4 & 10.3 & 12.7 & 7.1 \\
SF     & \textbf{1.5} & 11.5 & 14.7 & 21.3 & 23.9 \\
SF v2    & 4.7 & 10.1 & 14.3 & 16.7 & 22.7 \\
SF v2-stream    & 4.7 & 10.4 & 14.1 & 13.2 & 22.7 \\
pyannote    & 3.2 & 19.9 & 19.8 & 17.1 & 10.6 \\
pyannoteAI  & 2.7 & \textbf{9.9} & \textbf{9.1} & \textbf{10.1} & \textbf{6.6} \\
\bottomrule
\end{tabular}
\caption{DER (\%) aggregated by speaker count. For each column, we compute DER by averaging all samples with the corresponding number of speakers and averaging over them. Bold numbers indicate the best-performing system for each speaker count.}
\label{tab:der_by_spk}
\end{table}

\begin{table}[t!]
\centering
\begin{tabular}{lcccc|c}
\toprule
Model           & ALI  & AMI  & CH   & VC   & Avg   \\
\midrule
DiariZen        & 20.4 & 20.4 & 20.4 & 19.6 & 20.2\\
SF              & 166.7 & 160.7 & 177.9 & 153.4 & 164.7 \\
SF v2           & 223.6 & \textbf{223.7} & \textbf{212.5} & \textbf{197.4} & \textbf{214.3} \\
SF v2-stream    & \textbf{230.6} & 197.4 & \textbf{212.5} & \textbf{197.4} & 209.5\\
pyannote        & 45.5  & 47.6 & 43.5 & 43.5  & 45.0 \\
\bottomrule
\end{tabular}
\caption{Real-Time Factor (RTF, higher is better) computed as audio length divided by processing time. Processing time is only based on model inference time (without data or model loading). We do not report PyannoteAI's RTF, as we do not have access to model inference time through the API.}
\label{tab:rtf}
\end{table}

\section{Conclusion}
This work evaluated five state-of-the-art speaker diarization systems across four datasets containing a total of five languages.
Our evaluation found that PyannoteAI, a commercial closed-source model achieves the best overall performance, while DiariZen~\cite{han2024diarizen} provided competitive performance as an open-source alternative.
Sortformer v2 shows significant improvements over Sortformer, achieving exceptional computational efficiency at an average 214.3x real-time factor while maintaining comparable performance to DiariZen, particularly in low speaker-count scenarios.
Cross-lingual analysis indicates that annotated data scarcity for certain languages is a dominant driver in lower diarization performance, and analyzing failure cases, we found missed speech detection as the dominant failure mode across all models. Missed speech was found to be mainly caused by speech onset and end timing detection errors, highlighting possible future improvements for diarization models.

\clearpage

\bibliographystyle{IEEEbib}
\bibliography{refs}
\end{document}